\documentclass[%
superscriptaddress,
amsmath,amssymb,
aps,pra,
twocolumn,
floatfix,
]{revtex4}

\usepackage{graphicx}
\usepackage{dcolumn}
\usepackage{bm}
\usepackage{hyperref}

\usepackage{amsmath,amsfonts,amssymb,amsthm}
\usepackage{braket}
\usepackage{slashed}
\usepackage{comment}
\usepackage{float}
\usepackage{mhchem}
\usepackage{tabularx}
\def \be {\begin{equation}}
\def \ee {\end{equation}}
\usepackage{times}
\usepackage{nicefrac}
\usepackage{epsf}
\usepackage{bm}
\usepackage{bbm}
\usepackage{color}

\usepackage[labelformat=simple]{subcaption}

\DeclareCaptionLabelFormat{subcaptionlabel}{\normalfont(\textbf{#2}\normalfont)}
\preprint{APS/123-QED}
\begin{document}
\title{QED approach to valence-hole excitation in closed shell systems}
\author{R. N. Soguel}
\email{romain.soguel@uni-jena.de}
\affiliation{Theoretisch-Physikalisches Institut, Friedrich-Schiller-Universität Jena, Max-Wien-Platz 1, 07743 Jena, Germany}
\affiliation{Helmholtz-Institut Jena, Fr\"obelstieg 3, 07743 Jena, Germany}
\affiliation{GSI Helmholtzzentrum für Schwerionenforschung GmbH, Planckstraße 1, 64291 Darmstadt, Germany}
\author{A. V. Volotka}
\affiliation{School of Physics and Engineering, ITMO University, Kronverkskiy pr. 49, 197101 St. Petersburg, Russia}
\author{S. Fritzsche}
\affiliation{Theoretisch-Physikalisches Institut, Friedrich-Schiller-Universität Jena, Max-Wien-Platz 1, 07743 Jena, Germany}
\affiliation{Helmholtz-Institut Jena, Fr\"obelstieg 3, 07743 Jena, Germany}
\affiliation{GSI Helmholtzzentrum für Schwerionenforschung GmbH, Planckstraße 1, 64291 Darmstadt, Germany}
\date{\today}

\begin{abstract}
An \textit{ab initio} QED approach to treat a valence-hole excitation in closed shell systems is developed in the framework of the two-time-Green function method. The derivation considers a redefinition of the vacuum state and its excitation as a valence-hole pair. The proper two-time Green function, whose spectral representation confirms the poles at valence-hole excitation energies is proposed. An contour integral formula which connects the energy corrections and the Green function is also presented. First-order corrections to the valence-hole excitation energy involving self-energy, vacuum polarization, and one-photon-exchange terms are explicitly derived in the redefined vacuum picture. Reduction to the usual vacuum electron propagators is given that agrees in the Breit approximation with the many-body perturbation theory expressions for the valence-hole excitation energy.
\end{abstract}
\maketitle 

%
\section{Introduction}
Highly charged ions became a field of interest both from the theoretical and experimental sides. It has the great advantage to provide access to strong-field physics \cite{indelicato:2019:232001} and allow to probe quantum electrodynamics (QED) corrections up to the second-order in $\alpha$ (the fine structure constant) \cite{yerokhin:2008:062510, yerokhin:2010:57} although being a challenging task. Intensive experimental investigations have been carried over the years in a variety of system, ranging from H-like \cite{gumberidze:2005:223001, thorn:2009:163001, gassner:2018:073033}, He-like \cite{gumberidze:2004:203004, bruhns:2007:113001, Trassinelli_2009, amaro:2012:043005, chantler:2012:153001}, Be-like \cite{Feili_2005,Bernhardt_2015}, and Li-like \cite{brandau:2003:073202, beiersdorfer:2005:233003} to B-like \cite{draganic:2003:183001, mackel:2011:143002, liu:2021:062804} and F-like \cite{oneil:2020:032803, lu:2020:042817, wang:2022:arXiv} ions. Increasing experimental precision pushes theoretical predictions to their limits and enforces an accurate description of complex electron dynamics. Over the years, many approximated methods have been devised to access higher-order corrections, however, \textit{ab initio} calculations remain the holy grail in the quest for many-electron atoms in the frame of bound-state QED (BSQED).

Dealing with many-electron ions is a difficult task due to the numerical complexity involved as well as to derive formal BSQED expression. That is why ab initio calculations are limited so far to a few-electron ions \cite{artemyev:2005:062104, malyshev:2019:010501R, malyshev:2021:183001} and ions with single valence (or hole) electron \cite{kozhedub:2010:042513, sapirstein:2011:012504, sapirstein:2015:062508, volotka:2019:010502}. To facilitate the derivation of the formal expressions for many-electron systems the redefinition of the vacuum state is widely used in the relativistic many-body perturbation theory (MBPT) \cite{lindgren_morrison, johnson_2007}. However, within the BSQED it is not yet broadly employed. Previously, the vacuum redefinition method was employed within the BSQED mainly for single valence electron states \cite{shabaev:2002:119, volotka:2009:033005, glazov:2010:062112, volotka:2012:073001, volotka:2014:253004, Soguel_2021pra} and recently for two-valence electron states \cite{malyshev:2021:183001}. In Ref.~\cite{Soguel_2021pra} we showed that the employment of the redefined vacuum state allows one to keep track of one-electron gauge-invariant subsets into many-electron Feynman diagrams. Thus, several additional gauge-invariant subsets were identified. Later, we applied it to more sophisticated electron structures generalizing to either $N$ valence electrons or $N$ holes cases \cite{Soguel_2021sym}. As an example, we presented the complete set of formal expressions for BSQED corrections up to the second-order in $\alpha$ for the single-hole picture \cite{Soguel_2021sym}. Thus, the situation when both valence electrons and holes are involved in the description of a state has not been considered so far within the vacuum redefinition method.

The aim of this paper is to provide a rigorous \textit{ab initio} derivation of the BSQED perturbation theory for a valence-hole excitation in a closed shell system with the redefined vacuum approach. The two-time Green function (TTGF) formulation of the BSQED theory \cite{shabaev:2002:119} is employed as a mathematical tool for our derivation. The notion of a redefined vacuum state is used from the very beginning. It is shown that with the appropriate equal times choice conditions a Green function having the proper two-body state normalization in the non-interacting field limit can be constructed. Its spectral representation identifies poles at the valence-hole excitation energies and the integral formula for the energy correction to the binding energy is obtained. The latter expression is expanded to the first order, where one-particle radiative and one-photon exchange corrections are explicitly derived. Sec. \ref{section:BSQED} introduces the basics of BSQED and the redefinition of the vacuum state. The major part of the paper is devoted to the derivation of the TTGF suited for the valence-hole excitation energy, which is presented in Sec. \ref{section:Green}. Sec. \ref{section:1st_order} is dedicated to explicit deduction of the first-order corrections. Discussion and conclusion are found in Sec. \ref{section:discussion}. Some calculation details are provided in 
appendix \ref{appendix_B} for the zeroth-order TTGF.

Natural units ($\hbar = c = m_e = 1$) are used throughout this paper, the fine structure constant is defined as $\alpha = e^{2}/(4\pi),\,e < 0$. 
Unless explicitly stated, all integrals are meant to be on the interval  $\ ] -\infty, \infty \ [$.


\section{Bound state QED}
\label{section:BSQED}
The quantum relativistic description of the bound-state system under consideration relies on the Furry picture \cite{furry:1951:115} of QED, the so called bound-state QED. In this approach the eigenstates of the Dirac equation
\begin{equation} 
h_{\text{D}} \phi_j({\bf x}) = \left[-i \boldsymbol{\alpha} \cdot \boldsymbol{\nabla} + \beta  + V({\bf x})\right]\phi_j({\bf x})= \epsilon_j \phi_j({\bf x})\,,
\label{Dirac_eq}
\end{equation}
are solutions is the presence of an external classical Coulomb field arising from the nucleus, $V({\bf x}) = V_{\rm C}({\bf x})$. It means an all order treatment in $\alpha Z$, with $Z$ the nuclear charge, hence going beyond perturbative regime. The extended Furry picture implies the presence of a screening potential $U({\bf x})$ besides the Coulomb one, $V({\bf x}) = V_{\rm C}({\bf x}) + U({\bf x})$, which partially takes into account the interelectronic interaction. The time-dependent solution is obtained when $\phi_j({\bf x})$ is multiplied by the phase factor  $\exp{(-i\epsilon_j t)}$. $\alpha^{k}$ and $\beta$ are Dirac matrices and $j$ stands for all quantum numbers. The unperturbed normal ordered Hamiltonian is constructed as \cite{mohr:1998:227}
\begin{equation}
H_0 = \int d^{3}{\bf x} :\psi^{(0)\dagger}(t,{\bf x}) h_{\text{D}} \psi^{(0)}(t,{\bf x}):\,.
\end{equation}
A new vacuum state, named redefined vacuum state, is introduced in a way that all core orbitals from the closed shell belong to it \cite{lindgren_morrison}. It is denoted by $\ket{\alpha}$ notation,
\begin{equation}
\ket{\alpha} = a_{a}^{\dagger} a_{b}^{\dagger} ... \ket{0}\,.
\label{redefined vacuum}
\end{equation}
Here and further, we employ the MBPT notations of Lindgren and Morisson \cite{lindgren_morrison} and Johnson \cite{johnson_2007}: $v$ designates a valence electron,  $a,b,...$ stands for core orbitals, $i,j,...$ correspond to arbitrary states and $h$ to a hole. The redefinition of the vacuum state affects the non-interacting fermion field expansion in creation and annihilation operators and the electron propagator as such 
\begin{equation}
\psi^{(0)}_{\alpha}(t,{\bf x}) = \sum_{\epsilon_j > E_{\alpha}^{F} } a_j \phi_j({\bf x}) e^{-i\epsilon_j t} + \sum_{\epsilon_j < E_{\alpha}^{F} } b_j^{\dagger } \phi_j({\bf x}) e^{-i\epsilon_j t}
\label{psi_exp}
\end{equation}
and
\begin{eqnarray}
&&\bra{\alpha}  T\left[\psi_{\alpha}^{(0)}(t,{\bf x}) \bar{\psi}_{\alpha}^{(0)}(t^\prime,{\bf y})\right]\ket{\alpha} = \nonumber\\
&&= \frac{i}{2\pi} \int d\omega \sum_{j} \frac{\phi_j({\bf x}) \bar{\phi}_j({\bf y}) e^{-i(t - t^\prime) \omega}}{\omega - \epsilon_j + i\varepsilon (\epsilon_j-E_{\alpha}^{F}) }\,,
\label{eq:e_propagator}
\end{eqnarray}
respectively. The limit $\varepsilon \rightarrow 0$ is implied above, with $\varepsilon > 0$. $E_{\alpha}^{F}$ is the Fermi level of the redefined vacuum state lying slightly above the energy of the highest core state. 
%
%
%
%
We refer to Refs.~\cite{Soguel_2021pra, Soguel_2021sym} for more details on the vacuum redefinition within BSQED framework and its use in formula derivation. 

The interacting Hamiltonian takes a form
\begin{equation}
H_{\text{int}} = \int d^{3}{\bf x} :\psi^{(0)\dagger}(t,{\bf x}) h_{\text{int}} \psi^{(0)}(t,{\bf x}):\,,
\end{equation}
where $h_{\text{int}} = e \alpha^{\mu} A_{\mu}(t,{\bf x}) - U({\bf x})$ contains the interaction with the quantized electromagnetic field $A_{\mu}$ and the counterpotential term $-U({\bf x})$ when one works within the extended Furry picture. The interaction term is treated within BSQED perturbation theory. For its formulation there are several approaches \cite{mohr:1998:227, shabaev:2002:119, lindgren:2004:161, andreev:2008:135}. Our derivation presented in follows is based on the TTGF method \cite{shabaev:2002:119}.


\begin{widetext}

\section{Valence-hole Green function}
\label{section:Green}
Let us derive the Green function for the valence-hole excitation in a closed shell system and show that its spectral representation indeed has poles at valence-hole excitation energies. To begin with, we consider the general 4-point Green function
\begin{equation}
G(t_1^\prime, {\bf x}_1, t_2^\prime, {\bf x}_2, t_1, {\bf y}_1, t_2, {\bf y}_2) = \bra{0} T\left[\psi(t_1^{\prime},{\bf x}_1) \psi(t_2^{\prime}, {\bf x}_2) \bar{\psi}(t_2, {\bf y}_2) \bar{\psi}(t_1, {\bf y}_1)\right]\ket{0}\,.   
\end{equation}
This Green function contains all the information about the two-particle dynamics in presence of the nuclear Coulomb field. However, it is a difficult task to extract the necessary information. To get the energy levels it is enough to consider a two-time Green function. In the original work \cite{shabaev:2002:119}, Shabaev proposed to consider the following equal-time choice $t_1^{\prime} = t_2^{\prime} = t^{\prime}$ and $t_1 = t_2 = t$:
\begin{equation}
G(t^\prime, {\bf x}_1, t^\prime, {\bf x}_2, t, {\bf y}_1, t, {\bf y}_2) = \bra{0} T\left[\psi(t^{\prime},{\bf x}_1) \psi(t^{\prime}, {\bf x}_2) \bar{\psi}(t, {\bf y}_2) \bar{\psi}(t, {\bf y}_1)\right] \ket{0}\,.
\end{equation}
However, the spectral representation of the Green function for this particular choice of times unambiguously reveals poles only for pure electron (charge $2e$) or positron (charge $-2e$) states \cite{shabaev:2002:119}. It is a clear message that such an equal-time Green function can not deal with valence-hole excitation. Thus, one has to come up with a different Green function to describe such a system. Notice that although the choice of times was motivated and justified \textit{a posteriori} when the spectral representation is derived, it is nevertheless an arbitrary choice. \textit{A priori} one can also choose to have equal times such as $t_1^\prime = t_1 = t$ and $t_2^\prime = t_2 = t^\prime$, then the resulting Green function reads
\begin{equation}
G(t, {\bf x}_1, t', {\bf x}_2, t, {\bf y}_1, t', {\bf y}_2) = \bra{0} T\left[\psi(t,{\bf x}_1) \psi(t', {\bf x}_2) \bar{\psi}(t', {\bf y}_2) \bar{\psi}(t, {\bf y}_1)\right]\ket{0}\,.
\label{Green_initial}
\end{equation}
%
Similar Green functions were studied previously by Logunov and Tavkhelidze \cite{Logunov},  Fetter and Walecka \cite{fetter1971quantum}, Oddershede and J$\o$rgensen \cite{part-hole_Oddershede} and Liegener \cite{part_hole_Liegener}. In order to achieve the sough structure and for normalization reasons, as can be seen in appendix \ref{appendix_B}, one has to take into account three extra terms. One might argue that other structures are possible; in virtue of Ockham's razor the one proposed here is to our view the simplest one. Hence, the Green function one has to consider takes a form in the redefined vacuum $\ket{\alpha}$:
\begin{eqnarray}
&&G_{\alpha}(t_1, t_2; {\bf x}_1, {\bf x}_2, {\bf y}_1, {\bf y}_2) = \bra{\alpha} T\left[ \psi_\alpha(t_1,{\bf x}_1) \psi_\alpha(t_2, {\bf x}_2) \bar{\psi}_\alpha(t_2, {\bf y}_2) \bar{\psi}_\alpha(t_1, {\bf y}_1) - \psi_\alpha(t_1,{\bf x}_2) \psi_\alpha(t_2, {\bf x}_1) \bar{\psi}_\alpha(t_2, {\bf y}_2) \bar{\psi}_\alpha(t_1, {\bf y}_1) \right. \nonumber \\
&&-\left. \psi_\alpha(t_1,{\bf x}_1) \psi_\alpha(t_2, {\bf x}_2) \bar{\psi}_\alpha(t_2, {\bf y}_1) \bar{\psi}_\alpha(t_1, {\bf y}_2) + \psi_\alpha(t_1,{\bf x}_2) \psi_\alpha(t_2, {\bf x}_1) \bar{\psi}_\alpha(t_2, {\bf y}_1) \bar{\psi}_\alpha(t_1, {\bf y}_2) \right] \ket{\alpha} \,.    
\label{Green_vh}
\end{eqnarray}
To demonstrate that this Green function has the expected pole structure one has to consider its spectral representation, which is obtained by taking the Fourier transform of the Green function. For the sake of clarity, the steps to perform are briefly described. The first one is to rearrange the Dirac spinors to get identical times close to each other, having in mind that for equal time the only non-zero anti-commutator is  $\{\psi_\alpha(t,{\bf x}),\psi_\alpha^{\dagger}(t, {\bf y})\} = \delta^{(3)}({\bf x}-{\bf y})$. The next step is to proceed with the time ordering. Once it is done, a completeness relation $\mathbbm{1} = \sum_{\beta} \ket{\beta} \bra{\beta}$ is inserted to separate terms with different times within the time-ordered product. Then the Heisenberg representation for the field operator is introduced; $ \psi_\alpha(t, {\bf x})= e^{iHt} \psi_\alpha(0, {\bf x}) e^{-iHt}$ with $H = H_0 + H_\text{int}$ and the integral representation of the Heaviside function is applied. One ends up with
\begin{eqnarray}
&&\mathcal{G}_{\alpha}(E; {\bf x}_1, {\bf x}_2, {\bf y}_1, {\bf y}_2) \delta(E- E^{\prime}) = \frac{1}{2\pi i} \frac{1}{2!} \int dt_1 dt_2 e^{iEt_1 - i E^{\prime} t_2} G_{\alpha}(t_1, t_2; {\bf x}_1, {\bf x}_2, {\bf y}_1, {\bf y}_2) \nonumber \\&&= 
\frac{1}{4\pi^2} \frac{1}{2!} \int dt_1 dt_2 e^{iEt_1 - i E^{\prime} t_2} \int d\omega e^{-i\omega (t_1 - t_2)}  \left\{    \sum_{\beta}  \frac{ \mathcal{A} ({\bf x }_1, {\bf x}_2, {\bf y }_1, {\bf y }_2) }{\omega - E_{\beta} + i\varepsilon} 
- \sum_{\beta} \frac{  \mathcal{B} ({\bf x }_1, {\bf x}_2, {\bf y }_1, {\bf y }_2) }{\omega + E_{\beta} - i\varepsilon} \right\} \nonumber \\
&&= \frac{\delta(E- E^{\prime})}{2!}  \left\{  \sum_{\beta} \frac{ \mathcal{A}  ({\bf x }_1, {\bf x}_2, {\bf y }_1, {\bf y }_2) }{E - E_{\beta} + i\varepsilon} 
- \sum_{\beta} \frac{ \mathcal{B}  ({\bf x }_1, {\bf x}_2, {\bf y }_1, {\bf y }_2) }{E + E_{\beta} - i\varepsilon} \right\}\,,
\label{Fourier_Green}
\end{eqnarray}
where the $\mathcal{A}$ term is given by
\begin{eqnarray}
&&\mathcal{A}({\bf x }_1, {\bf x}_2, {\bf y }_1, {\bf y }_2) = \bra{\alpha} \left[ \psi_{\alpha}(0,{\bf x}_1)  \bar{\psi}_{\alpha}(0, {\bf y}_1) \ket{\beta} \bra{\beta} \psi_{\alpha}(0, {\bf x}_2) \bar{\psi}_{\alpha}(0,{\bf y}_2)
- \psi_{\alpha}(0,{\bf x}_2) \bar{\psi}_{\alpha}(0, {\bf y}_1) \ket{\beta} \bra{\beta} \psi_{\alpha}(0, {\bf x}_1) \bar{\psi}_{\alpha}(0, {\bf y}_2)  \right. \nonumber \\
&&-\left. \psi_{\alpha}(0,{\bf x}_1) \bar{\psi}_{\alpha}(0, {\bf y}_2) \ket{\beta} \bra{\beta} \psi_{\alpha}(0, {\bf x}_2) \bar{\psi}_{\alpha}(0, {\bf y}_1)  
+ \psi_{\alpha}(0,{\bf x}_2) \bar{\psi}_{\alpha}(0, {\bf y}_2) \ket{\beta} \bra{\beta} \psi_{\alpha}(0, {\bf x}_1) \bar{\psi}_{\alpha}(0, {\bf y}_1)  \right] \ket{\alpha}\,,
\label{a_term}
\end{eqnarray}
and the $\mathcal{B}$ one is 
\begin{eqnarray}
&&\mathcal{B}  ({\bf x }_1, {\bf x}_2, {\bf y }_1, {\bf y }_2) = \bra{\alpha} \left[ \psi_{\alpha}(0, {\bf x}_2) \bar{\psi}_{\alpha}(0,{\bf y}_2) \ket{\beta} \bra{\beta}  \psi_{\alpha}(0,{\bf x}_1)  \bar{\psi}_{\alpha}(0, {\bf y}_1)
-  \psi_{\alpha}(0, {\bf x}_1) \bar{\psi}_{\alpha}(0, {\bf y}_2) \ket{\beta} \bra{\beta} \psi_{\alpha}(0,{\bf x}_2) \bar{\psi}_{\alpha}(0, {\bf y}_1) \right. \nonumber \\
&&- \left. \psi_{\alpha}(0, {\bf x}_2) \bar{\psi}_{\alpha}(0, {\bf y}_1)  \ket{\beta} \bra{\beta} \psi_{\alpha}(0,{\bf x}_1) \bar{\psi}_{\alpha}(0, {\bf y}_2) 
+  \psi_{\alpha}(0, {\bf x}_1) \bar{\psi}_{\alpha}(0, {\bf y}_1)  \ket{\beta} \bra{\beta} \psi_{\alpha}(0,{\bf x}_2) \bar{\psi}_{\alpha}(0, {\bf y}_2) \right] \ket{\alpha}\,.
\end{eqnarray}
Thus, we derived the spectral representation of expression (\ref{Green_vh}). Under the assumption of non-interacting electron-positron fields and their expansion in creation and annihilation operators, as in Eq.~(\ref{psi_exp}), the only consistent zeroth-order $\ket{\beta}$ states are found to be 
\begin{equation}
\ket{\beta} = \left\{ \ket{vh}=  a_v^{\dagger} b_h^{\dagger} \ket{\alpha}, \,\,  \ket{\alpha} \right\}\,.
\end{equation}
Now that the Green function spectral representation is obtained as a function of $E$ one can define its analytic continuation in the complex $E$ plane. Then, one sees the presence of poles at the valence-hole excitation energies $E_{vh}$ and $-E_{vh}$ as well as at the zero (vacuum energy). Some remarks are important to be noticed here. First, despite Eq.~(\ref{Fourier_Green}) looks similar to the one obtained in Ref.~\cite{shabaev:2002:119} it has poles at essentially different energies. Second, although the structure of $\mathcal{A}$ and $\mathcal{B}$ looks complicated, in the non-interacting cases it contains neutral charged states corresponding to valence-hole excitations of a closed shell. Third and most important, it leads to normalized two-particle wavefunctions in the zeroth order, as can be seen in appendix \ref{appendix_B}. A coordinate integrated Green function is built out of spectral representation of the Green in the following manner
\begin{equation}
g_{\alpha}(E) =\frac{1}{2!} \int d^3 {\bf x}_1 d^3 {\bf x}_2 d^3 {\bf y}_1 d^3 {\bf y}_2 \, : \psi_{\alpha}^{(0)\dagger}({\bf x}_1) \psi_{\alpha}^{(0)\dagger}({\bf x}_2) \mathcal{G}_{\alpha}(E; {\bf x}_1, {\bf x}_2, {\bf y}_1, {\bf y}_2 ) \gamma^{0}_1 \gamma^{0}_2 \psi_{\alpha}^{(0)}({\bf y}_2) \psi_{\alpha}^{(0)}({\bf y}_1): \,.
\label{Green}
\end{equation}
Further, we employ the occupation number representation as in MBPT description provided by Lindgren \cite{lindgren} to construct the 2-particle operator. Since our interest lies in valence-hole state described by \cite{avgoustoglou1992many, PhysRevA.51.297}
\begin{equation}
    \ket{(vh)_{JM}} = \sum_{m_v, m_h} \braket{j_v m_v j_h -m_h|JM}(-1)^{j_h - m_h} a_v^{\dagger} b_h^{\dagger} \ket{\alpha} \equiv F_{vh}  a_v^{\dagger} b_h^{\dagger} \ket{\alpha}\,,
    \label{vh_state}
\end{equation}
where the $jj$-coupling scheme is applied to form a state with total angular momentum $J$ and its projection $M$, one works out and retains only the six terms involving two $a$'s and two $b$'s operators. After normal ordering one arrives to the expression
\begin{eqnarray}
g_{\alpha}(E) &\cong& \frac{1}{2!} \left\{
\sum_{i,j>E_{\alpha}^F, k,l<E_{\alpha}^F}  a_i^{\dagger} a_j^{\dagger} b_l^{\dagger} b_k^{\dagger}
- \sum_{k,l>E_{\alpha}^F, i,j<E_{\alpha}^F}  a_k a_l b_i b_j
+ \sum_{i,l>E_{\alpha}^F, j,k<E_{\alpha}^F}  a_i^{\dagger} a_l b_k^{\dagger} b_j  \right.\nonumber \\
&+& \left. \sum_{j,k>E_{\alpha}^F, i,l<E_{\alpha}^F}  a_j^{\dagger} a_k b_l^{\dagger} b_i
- \sum_{i,k>E_{\alpha}^F, j,l<E_{\alpha}^F}  a_i^{\dagger} a_k b_l^{\dagger} b_j
- \sum_{j,l>E_{\alpha}^F, i,k<E_{\alpha}^F}  a_j^{\dagger} a_l b_k^{\dagger} b_i
\right\}g_{\alpha, ijkl} (E) \,,
\label{G_operator}
\end{eqnarray}
with
\begin{eqnarray}
g_{\alpha, ijkl}(E) = \int d^3{\bf x}_1 d^3{\bf x}_2 d^3{\bf y}_1 d^3{\bf y}_2 \phi^{\dagger}_i({\bf x}_1) \phi^{\dagger}_j({\bf x}_2)  \mathcal{G}_{\alpha}(E; {\bf x}_1, {\bf x}_2, {\bf y}_1, {\bf y}_2 )  \gamma^{0}_1 \gamma^{0}_2 \phi_k({\bf y}_1) \phi_l({\bf y}_2)\,.
\label{G_matrix_element}
\end{eqnarray}
Further, we have to evaluate the matrix element of $g_\alpha(E)$ with the valence-hole state defined by Eq.~(\ref{vh_state}). The first two terms in Eq.~(\ref{G_operator}) do not contribute since they can not be fully contracted with the valence-hole state. Computing the matrix element one gets
\begin{eqnarray}
 \bra{(v h)_{JM}} g_{\alpha}(E) \ket{(v h)_{JM} } =  F_{v_1 h_1} F_{v_2 h_2} \left[ g_{\alpha, v_1 h_2 h_1 v_2}(E) - g_{\alpha, v_1 h_2 v_2 h_1}(E)\right]\,.
\label{G_matrix}
\end{eqnarray}
Now all the necessary pieces are available to derive the energy correction formula to the valence-hole binding energy. Applying the integral formalism developed in Ref.~\cite{shabaev:2002:119} and focusing only on the first term  with the contour $\Gamma_{vh}$ surrounding only the pole
$E \sim E_{vh}^{(0)}$,
\begin{equation}
E_{vh}^{(0)} = \bra{vh} H_0 \ket{vh} = \epsilon_v - \epsilon_h\,,
\label{eq:0-rv}
\end{equation}
one ends up to the expression:
\begin{equation}
E_{vh} = \frac{ \displaystyle{ \frac{1}{2\pi i} \oint_{\Gamma_{vh}} dE E \bra{ (vh)_{JM} } g_{\alpha} (E)\ket{ (vh)_{JM} } }}{ \displaystyle{ \frac{1}{2\pi i} \oint_{\Gamma_{vh}} dE \bra{ (vh)_{JM} } g_{\alpha} (E)\ket{ (vh)_{JM} }}} \,.
\end{equation}
Evaluating the zeroth order Green function 
\begin{equation}
\bra{(vh)_{JM} } g_{\alpha}^{(0)} (E)\ket{ (vh)_{JM} } = \frac{1}{E- E_{vh}^{(0)}} + \text{regular terms at } E \sim E_{vh}^{(0)}
\label{Green_zero}
\end{equation}
(detailed calculation is presented in appendix \ref{appendix_B}), we derive also the expression for the energy shift $\Delta E_{vh} = E_{vh} - E_{vh}^{(0)}$:
\begin{equation}
\Delta E_{vh} = \frac{ \displaystyle{ \frac{1}{2\pi i} \oint_{\Gamma_{vh}} dE (E - E_{vh}^{(0)} ) \bra{ (vh)_{JM} } \Delta g_{\alpha} (E)\ket{ (vh)_{JM} } }}{ \displaystyle{ 1 + \frac{1}{2\pi i} \oint_{\Gamma_{vh}} dE  \bra{ (vh)_{JM} } \Delta g_{\alpha} (E)\ket{ (vh)_{JM} } }} \,,
\label{eq:energy_corr}
\end{equation}
where $\Delta g_{\alpha}(E) = g_{\alpha}(E) - g^{(0)}_{\alpha}(E)$. Expanding $\Delta g_{\alpha}(E)$ into a series in $\alpha$, $\Delta g_{\alpha}(E) = \Delta g^{(1)}_{\alpha}(E) + \Delta g^{(2)}_{\alpha}(E) + ...$ and combining the terms of the same order one easily obtains the BSQED perturbation expansion for the valence-hole energy  $\Delta E_{vh} = \Delta E^{(1)}_{vh} + \Delta E^{(2)}_{vh} + ...$. Thus, in this section we obtain the two-time Green function suited for the treatment of valence-hole states and derive the formula expressing the energy corrections as contour integrals of the Green function. In the next section we apply this formalism for the evaluation of the first-order corrections to the energy of the valence-hole excitation.
%

\section{First-order corrections}
\label{section:1st_order}
The first-order energy correction, obtained from the expansion of Eq.~(\ref{eq:energy_corr}), yields
\begin{equation}
\Delta E_{vh}^{(1)} =   \frac{1}{2\pi i} F_{v_1 h_1} F_{v_2 h_2}\oint_{\Gamma_{vh}} dE (E - E_{vh}^{(0)} )  \left[ \Delta g^{(1)}_{\alpha, v_1 h_2 h_1 v_2}(E) - \Delta g^{(1)}_{\alpha, v_1 h_2 v_2 h_1}(E)\right]\,.
\label{eq:1st_order}
\end{equation}
\end{widetext}
The energy correction can be split into  one- $\Delta E_{vh}^{(1)1}$ and two-particle $\Delta E_{vh}^{(1)2}$ terms:
\begin{equation}
\Delta E_{vh}^{(1)} = \Delta E_{vh}^{(1)1} + \Delta E_{vh}^{(1)2}\,.
\end{equation}
In turn, the one-particle contribution originates from three Feynman diagrams: self energy (SE), vacuum polarization (VP), and counterpotential (CP) depicted in Fig.~\ref{fig:1_1}(a), which give rise to the following terms in the Green function:
\begin{equation}
\Delta g_{\alpha}^{(1)1}(E) = \Delta g_{\alpha}^{(1)\text{SE}}(E) + \Delta g_{\alpha}^{(1)\text{VP}}(E) + \Delta g_{\alpha}^{(1)\text{CP}}(E)\,.
\end{equation}
The two-particle contribution corresponds to the Green function $\Delta g_{\alpha}^{(1)2}$ and the valence-hole one-photon exchange diagram presented in Fig.~\ref{fig:1_1}(b). 
\begin{figure}
\begin{subfigure}[b]{0.39\textwidth}
         \includegraphics[width=\textwidth]{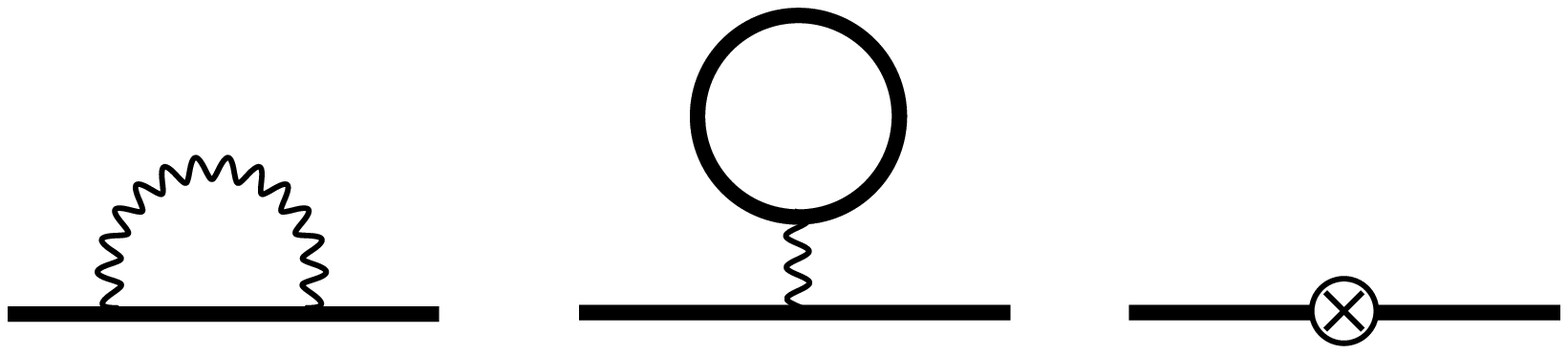}
         \caption{}
        \label{fig:11}
     \end{subfigure}
     \hfill
     \begin{subfigure}[b]{0.06\textwidth}
        \includegraphics[width=\textwidth]{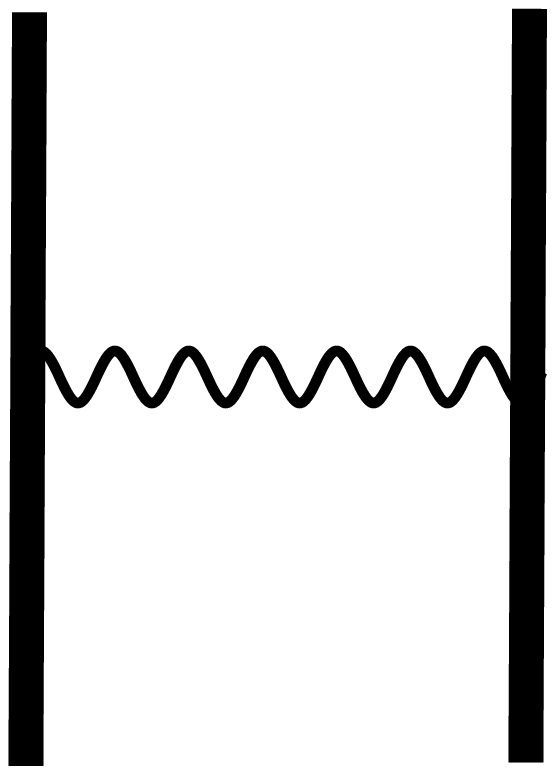}
        \caption{}
        \label{fig:12}
     \end{subfigure}
\caption{(a) the first-order one-particle Feynman diagrams corresponding, from left to right, to the self-energy (SE), vacuum-polarization (VP), and counterpotential (CP) contributions and (b) the valence-hole one-photon exchange Feynman diagram. Single solid lines indicate the electron propagators in the redefined vacuum representation. Wavy lines correspond to the photon propagator and the cross inside a circle represents a counterpotential term, $-U({\bf x})$.}
\label{fig:1_1}
\end{figure}
We start the consideration with the one-particle diagrams.
%
%
\subsection{One-particle contributions}
Formally, for single particle graphs, SE, VP or CP, a second disconnected line (propagator) is present. However, these diagrams reduce to the single-particle case since the disconnected line can be integrated out, as we will show it below. In what follows we consider in details the SE correction, which can be further divided into two terms, one where the SE loop is located on the valence electron line $\Delta g^{(1) \text{SE}v}_{\alpha}$ and another one on the hole line $\Delta g^{(1) \text{SE}h}_{\alpha}$. Let us consider the valence SE graph with disconnected hole line first. The Feynman rules, according to \cite{shabaev:2002:119} but noticing that hole's energy formally flows in the negative $t$ direction while being a positive quantity, provides us with the following expression:
\begin{widetext}
\begin{eqnarray}
\Delta g^{(1) \text{SE}v}_{\alpha, v_1 h_2 v_2 h_1} (E) \delta(E- E^{\prime}) &=& e^2 \left( \frac{i}{2\pi}\right)^2 \int d^3{\bf x} d^3{\bf y} dp_1^0 dp_2^0 d\omega dk^0 \delta (E- p_1^0 + p_2^0) \delta (E^{\prime} - p_1^{0} + p_2^0) \delta (p_1^0 - \omega - k^0) \nonumber \\
&\times& \frac{\bar{\psi}_{v_1} ({\bf y })}{p_1^0 - \epsilon_{v} + i \varepsilon} \gamma^{\mu}  \sum_j \frac{ \psi_j ({\bf y })  \bar{\psi}_j ({\bf x})}{k^0 - \epsilon_j + i \varepsilon(\epsilon_j - E_{\alpha}^F)}  D_{\mu \nu} (\omega, {\bf y} - {\bf x}) \gamma^{\nu} \frac{\psi_{v_2} ({\bf x})}{p_1^0 - \epsilon_{v} + i \varepsilon} \frac{ \delta_{h_1 h_2 }}{p_2^0 - \epsilon_{h} - i\varepsilon} \nonumber \\
&=& \left( \frac{i}{2\pi}\right)^2 \int dp_1^0 d\omega \sum_j \frac{  I_{v_1 j j v_2 }(\omega)}{ p_1^0 - \omega - \epsilon_j + i \varepsilon(\epsilon_j - E_{\alpha}^F)}  \frac{\delta (E - E^{\prime})}{\left[p_1^0 - \epsilon_{v} + i \varepsilon\right]^2} \frac{ \delta_{h_1 h_2 }}{p_1^0 - E - \epsilon_{h} -i\varepsilon}\,,
\end{eqnarray}
where two momentum integration have been carried out with the help of delta functions, and further simplified due to the orthogonality of the wavefunctions. We highlight the fact that  $p_2^0$ and $p_2^{\prime 0}$ are the hole's energy in all calculations and that the delta functions taking care of the energy conservation of initial and final states are affected as can be expected from eq.~(\ref{eq:0-rv}). Notice also that the zeroth-order energy of each particle does not depend on its spin projection, i.e., $\epsilon_{v_1} = \epsilon_{v_2} = \epsilon_{v}$, $\epsilon_{h_1} = \epsilon_{h_2} = \epsilon_{h}$. The interelectronic-interaction matrix element $I_{ijkl}(\omega)$ is a shorthand notation standing for
\begin{equation}
I_{i j k l}(\omega) = \int d^3{\bf x} d^3{\bf y} \psi_i^{\dagger}({\bf x}) \psi_j^{\dagger}({\bf y}) I({\bf x} - {\bf y};\omega)  \psi_k({\bf x}) \psi_l({\bf y})\,,
\end{equation}
and satisfies the transposition symmetry property $I_{i j k l}(\omega) = I_{j i l k}(\omega)$. The interelectronic-interaction operator $I({\bf x}- {\bf y};\omega)$ is defined as $I({\bf x}- {\bf y}; \omega) = e^2 \alpha^{\mu} \alpha^{\nu} D_{\mu \nu} ({\bf x}- {\bf y}; \omega)$ where $\alpha^{\mu}  = (1, \boldsymbol{\alpha})$,  $D_{\mu \nu} ({\bf x} - {\bf y};\omega)$ is the photon propagator and $\omega$ is the photon's energy. Since the integration over energy $E$ is the next step, the singularities in $E - E_{vh}^{(0)}$ should be analyzed. Notice that %
\begin{equation}
\frac{1}{\left[p_1^0 - \epsilon_{v} + i\varepsilon\right]^2} \frac{1}{p_1^0 - E - \epsilon_{h}  -i\varepsilon}  =
\frac{1}{(E - E_{v h}^{(0)})^2} \left[\frac{1}{ p_1^0 - E - \epsilon_{h}  -i\varepsilon} - \frac{1}{p_1^0 - \epsilon_{v} + i\varepsilon } \right] -
\frac{1}{E - E_{v h}^{(0)} } \frac{1}{\left[p_1^0 - \epsilon_{v} + i\varepsilon\right]^2}\,. 
\end{equation}
Only the most singular part is to retain. It leads to 
\begin{eqnarray}
\Delta E_{vh}^{(1) \text{SE}v} &=& -\frac{1}{2\pi i}F_{v_1 h_1} F_{v_2 h_2} \oint_{\Gamma_{vh}} dE (E - E_{vh}^{(0)} ) \Delta g^{(1) \text{SE}v}_{\alpha, v_1 h_2 v_2 h_1} (E)  \nonumber \\
&=& \frac{i}{2\pi} F_{v_1 h_1} F_{v_2 h_2} \int d\omega  \sum_j \frac{  I_{v_1 j j v_2 }(\omega) \delta_{h_1 h_2 } }{ \epsilon_v - \omega - \epsilon_j + i \varepsilon(\epsilon_j - E_{\alpha}^F)}  \equiv F_{v_1 h_1} F_{v_2 h_2}  \delta_{h_1 h_2 } \bra{v_1}  \Sigma_{\alpha}(\epsilon_{v}) \ket{v_2} \,,
\label{eq:se}
\end{eqnarray}
where the $\Sigma_\alpha$ stands for the the SE operator in the redefined vacuum framework and the identity:
\begin{eqnarray}
\delta(x) = \frac{i}{2\pi} \left( \frac{1}{x + i \varepsilon} + \frac{1}{-x + i \varepsilon} \right)
\end{eqnarray}
was utilized. Since the investigated graph is the SE matrix element for a single valence electron, there is no need to distinguish between the initial state $v_1$ and the final state $v_2$ because the SE operator preserves the spin projection, which was the solely difference between them. Hence one can write $\delta_{v_1 v_2}$ and the previous prefactor is then $F_{v_1 h_1} F_{v_2 h_2} \delta_{v_1 v_2} \delta_{h_1 h_2} = 1$ \cite{avgoustoglou1992many}. Same calculation is performed for the SE graph locating on the hole line and the electron line is integrated out this time:
\begin{equation}
\Delta g^{(1) \text{SE}h}_{\alpha, v_1 h_2 v_2 h_1} (E) =  \left( \frac{i}{2\pi}\right)^2 \int dp_2^0 d\omega \sum_j \frac{  I_{h_2 j j h_1 }(\omega)  }{ p_2^0 - \omega - \epsilon_j  + i \varepsilon(\epsilon_j - E_{\alpha}^F) }  \frac{1}{\left[p_2^0 - \epsilon_{h} - i\varepsilon\right]^2} \frac{ \delta_{v_1 v_2 }}{E + p_2^0- \epsilon_{v} + i\varepsilon} \,.
\end{equation}
As previously one isolates the most singular part 
\begin{equation}
\frac{1}{\left[p_2^0 - \epsilon_{h} - i \varepsilon\right]^2} \frac{1}{E + p_2^0- \epsilon_{v} + i\varepsilon} = \frac{1}{E - E_{v h}^{(0)} } \frac{1}{\left[p_2^0 - \epsilon_{h} - i \varepsilon\right]^2}
-  \frac{1}{(E - E_{v h}^{(0)})^2  } \left[ \frac{1}{ p_2^0  - \epsilon_{h}  -i\varepsilon} - \frac{1}{E + p_2^0 - \epsilon_{v} + i\varepsilon } \right]\,.
\end{equation}
Thus, one gets 
\begin{eqnarray}
\Delta E_{vh}^{(1) \text{SE}h} &=& -\frac{1 }{2\pi i} F_{v_1 h_1} F_{v_2 h_2} \oint_{\Gamma_{vh}} dE (E - E_{vh}^{(0)} ) \Delta g^{(1) \text{SE}h}_{\alpha, v_1 h_2 v_2 h_1} (E)  \nonumber \\
&=&- \frac{i}{2\pi} F_{v_1 h_1} F_{v_2 h_2}  \int d\omega  \sum_j \frac{  I_{h_2 j j h_1 }(\omega) \delta_{v_1 v_2 }  }{ \epsilon_{h_1} - \omega - \epsilon_j + i \varepsilon(\epsilon_j - E_{\alpha}^F)} \equiv -F_{v_1 h_1} F_{v_2 h_2}  \delta_{v_1 v_2 } \bra{h_2} \Sigma_{\alpha}(\epsilon_{h}) \ket{h_1} \,.
\end{eqnarray}
As before, the SE contribution to a single hole state is considered, thus $\delta_{h_1 h_2}$ and the prefactor reduces to $1$. The one-particle result is recovered in both cases. Such calculations can be extended to VP graph under the modification $\bra{v_1}\Sigma_{\alpha}(\epsilon_{v}) \ket{v_2}$ to
\begin{eqnarray}
\bra{v_1} \Upsilon_{\alpha} \ket{v_2}    &=& -\frac{ie^2}{2\pi} \int d^3{\bf x} d^3{\bf y} \psi_{v_1}^{\dagger} ( { \bf y }) \alpha^{\mu} D_{\mu \nu }(0, { \bf y } - { \bf x }) \int d\omega \text{Tr} \left[ \sum_j \frac{ \psi_j ( { \bf x }) \psi^{\dagger}_j ( { \bf x })  }{ \omega - \epsilon_j + i \varepsilon(\epsilon_j - E_{\alpha}^F)} \alpha^{\nu} \right] \psi_{v_2} ( { \bf y }) \nonumber \\ 
&\equiv& -\frac{i}{2\pi} \int d\omega \sum_j \frac{ I_{v_1 j v_2 j}(0) }{ \omega - \epsilon_j + i \varepsilon(\epsilon_j - E_{\alpha}^F)}
\label{eq:vp}
\end{eqnarray}
and accordingly for the hole case: $\bra{h_2}\Sigma_{\alpha}(\epsilon_{h}) \ket{h_1}$ to $\bra{h_2}\Upsilon_{\alpha} \ket{h_1}$. Hence, the one-particle graph contributions in the redefined vacuum framework are given by 
\begin{eqnarray}
\Delta E_{vh}^{(1)1} &=&
F_{v_1 h_1} F_{v_2 h_2} \delta_{v_1 v_2 } \delta_{h_1 h_2} \left[ \bra{v_1} \Sigma_{\alpha}(\epsilon_{v}) \ket{v_2}  +\bra{v_1} \Upsilon_{\alpha} \ket{v_2} -U_{v_1 v_2} - \bra{h_2} \Sigma_{\alpha}(\epsilon_{h}) \ket{h_1} - \bra{h_2} \Upsilon_{\alpha} \ket{h_1} + U_{h_2 h_1}\right] \nonumber\\
&=&  \bra{v} \Sigma_{\alpha}(\epsilon_{v}) \ket{v}  +\bra{v} \Upsilon_{\alpha} \ket{v} -U_{v v} - \bra{h} \Sigma_{\alpha}(\epsilon_{h}) \ket{h} - \bra{h} \Upsilon_{\alpha} \ket{h} + U_{h h} \,.
\label{eq:1-part}
\end{eqnarray}
Two counterpotential terms $U_{ij}= \bra{i} U \ket{j}$ are added into the formula above to accommodate for the extended Furry picture. Their treatment is straightforward and does not need to be given in details.

%
\subsection{Two-particle contributions}
The only two-particle correction found at this order is the valence-hole one-photon exchange, exchange part $\Delta E_{vh}^{(1)2\text{exc}}$ minus direct part $\Delta E_{vh}^{(1)2\text{dir}}$ according to Eq.~(\ref{eq:1st_order}):
\begin{eqnarray}
\Delta E_{vh}^{(1)2\text{exc}} &=& \frac{1}{2\pi i} F_{v_1 h_1} F_{v_2 h_2}\oint_{\Gamma_{vh}} dE (E - E_{vh}^{(0)} )  \Delta g^{(1)2\text{exc}}_{\alpha, v_1 h_2 h_1 v_2}(E)\,,\\
\Delta E_{vh}^{(1)2\text{dir}} &=& -\frac{1}{2\pi i} F_{v_1 h_1} F_{v_2 h_2}\oint_{\Gamma_{vh}} dE (E - E_{vh}^{(0)} )  \Delta g^{(1)2\text{dir}}_{\alpha, v_1 h_2 v_2 h_1}(E)\,.
\end{eqnarray}
Let us tackle first the direct graph keeping in mind that the hole's energies are flowing backward in time. Similar to previous calculations, trivial steps are already performed and the expression is given by
\begin{eqnarray}
\Delta g^{(1)\text{dir}}_{\alpha, v_1 h_2 v_2 h_1} (E) \delta (E- E^{\prime}) &=& e^2\left( \frac{i}{2\pi}\right)^2 \int d^3{\bf x} d^3 {\bf y} dp_1^0 dp_1^{\prime 0} dp_2^0 dp_2^{\prime 0} d\omega \delta (E- p_1^0 +p_2^{0}) \delta (E^{\prime} - p_1^{\prime 0} + p_2^{\prime 0}) \delta (p_1^{0} - \omega - p_1^{\prime 0}) \nonumber \\
&\times&  \delta (p_2^{\prime 0} + \omega - p_2^{0} ) \frac{\bar{\psi}_{v_1} ({\bf x})}{ p_1^{\prime 0} - \epsilon_{v} + i\varepsilon }  \frac{\bar{\psi}_{h_2} ({\bf y})}{ p_2^{\prime 0} - \epsilon_{h} - i\varepsilon } \gamma^{\mu } \gamma^{\nu } D_{\mu \nu } (\omega, { \bf x } - { \bf y}) \frac{\psi_{v_2} ({\bf x}) }{ p_1^0 - \epsilon_{v} + i \varepsilon} \frac{\psi_{h_1} ({\bf y}) }{ p_2^0 - \epsilon_{h} - i \varepsilon}\nonumber \\
&=& \left( \frac{i}{2\pi}\right)^2 \int dp_2^0 dp_2^{\prime 0} \frac{I_{v_1 h_2 v_2 h_1}(p_2^0 -p_2^{\prime 0})}{E+ p_2^{\prime 0} - \epsilon_{v} + i\varepsilon }  \frac{1}{ p_2^{\prime 0} - \epsilon_{h} - i\varepsilon }\frac{\delta (E - E^{\prime})}{ E + p_2^0 - \epsilon_{v} + i \varepsilon}  \frac{1 }{ p_2^0 - \epsilon_{h} - i \varepsilon} \,.
\end{eqnarray}
Rewriting the denominators to pull out the singular part as
\begin{eqnarray}
\frac{1}{E+ p_2^{\prime 0} - \epsilon_{v} + i\varepsilon }  \frac{1}{ p_2^{\prime 0} - \epsilon_{h} - i\varepsilon } &=& \frac{1}{E- E^{(0)}_{v h}} \left(  \frac{1}{ p_2^{\prime 0} - \epsilon_{h} - i\varepsilon } -  \frac{1}{E+ p_2^{\prime 0} - \epsilon_{v} + i\varepsilon } \right) \,, \nonumber \\ 
\frac{1 }{ E + p_2^0 - \epsilon_{v} + i \varepsilon}  \frac{1 }{ p_2^0 - \epsilon_{h} - i \varepsilon} &=& \frac{1}{E-  E^{(0)}_{v h}} \left(  \frac{1 }{ p_2^0 - \epsilon_{h} - i \varepsilon} - \frac{1 }{ E + p_2^0 - \epsilon_{v} + i \varepsilon}  \right)\,,
\end{eqnarray}
we get to
\begin{eqnarray}
\Delta E_{vh}^{(1)2\text{dir}} = 
- F_{v_1 h_1} F_{v_2 h_2} I_{v_1 h_2 v_2 h_1} (0) \,.
\end{eqnarray}
Last but not least is the exchange graph
. The partially simplified expression is found to be
\begin{eqnarray}
\Delta g^{(1)2\text{exc}}_{\alpha, v_1 h_2 h_1 v_2} (E) \delta (E- E^{\prime}) &=& e^2\left( \frac{i}{2\pi}\right)^2 \int d^3{\bf x} d^3{\bf y} dp_1^0 dp_1^{\prime 0} dp_2^0 dp_2^{\prime 0} d\omega \delta (E- p_1^0 +p_2^{0}) \delta (E^{\prime} - p_1^{\prime 0} + p_2^{\prime 0}) \delta (p_2^{0} - \omega - p_1^{\prime 0}) \nonumber \\
&\times&\delta (p_2^{\prime 0} + \omega - p_1^{0}) \frac{\bar{\psi}_{v_1} ({\bf x})}{ p_1^{\prime 0} - \epsilon_{v} + i\varepsilon }  \frac{\bar{\psi}_{h_2} ({\bf y})}{ p_2^{\prime 0} - \epsilon_{h} - i\varepsilon }\gamma^{\mu } \gamma^{\nu } D_{\mu \nu } (\omega, { \bf x} - { \bf y}) \frac{\psi_{h_1} ({\bf x}) }{ p_2^0 - \epsilon_{h} - i \varepsilon}
\frac{\psi_{v_2} ({\bf y}) }{ p_1^0 - \epsilon_{v} + i \varepsilon} \nonumber \\
&=& \left( \frac{i}{2\pi}\right)^2 \int dp_1^0 dp_2^{\prime 0} \frac{I_{v_1 h_2 h_1 v_2} (p_2^{\prime 0} - p_1^0)}{E+ p_2^{\prime 0} - \epsilon_{v} + i\varepsilon }  \frac{1}{ p_2^{\prime 0} - \epsilon_{h} - i\varepsilon }\frac{\delta (E - E^{\prime})}{ p_1^0 - E - \epsilon_{h} - i \varepsilon}  \frac{1 }{ p_1^0 - \epsilon_{v} + i \varepsilon} \,.
\end{eqnarray}
As before the singular part of the denominators are separated. Hence, the energy integration gives us
\begin{eqnarray}
\Delta E_{vh}^{(1)2\text{exc}} =  
  F_{v_1 h_1} F_{v_2 h_2} I_{v_1 h_2 h_1 v_2} (\Delta_{h v})\,,
\end{eqnarray}
here we introduced $\Delta_{h v} = \epsilon_{h} - \epsilon_{v}$, and the total two-particle contribution is 
\begin{eqnarray}
\Delta E_{vh}^{(1)2} =  F_{v_1 h_1} F_{v_2 h_2} \left[ I_{v_1 h_2 h_1 v_2} (\Delta_{h v}) - I_{v_1 h_2 v_2 h_1} (0)  \right] \,.
\label{eq:2-part}
\end{eqnarray}
\end{widetext}

%
\subsection{Usual vacuum description}
So far the formulas for the first-order corrections obtained in the previous subsections,
\begin{eqnarray}
\Delta E_{vh}^{(1)} &=&
F_{v_1 h_1} F_{v_2 h_2} \left[ I_{v_1 h_2 h_1 v_2} (\Delta_{h v}) - I_{v_1 h_2 v_2 h_1} (0)\right]\nonumber\\
&+&\bra{v} \Sigma_{\alpha}(\epsilon_{v}) \ket{v}  +\bra{v} \Upsilon_{\alpha} \ket{v} -U_{v v}\nonumber\\
&-&\bra{h} \Sigma_{\alpha}(\epsilon_{h}) \ket{h} - \bra{h} \Upsilon_{\alpha} \ket{h} + U_{h h}\,,
\label{eq:1-rv}
\end{eqnarray}
are written for the case when the redefined vacuum state is employed in the electron propagator, see Eq.~(\ref{eq:e_propagator}). We use also $\alpha$ subscript in $\Sigma_\alpha$ and $\Upsilon_\alpha$ operators to emphasize this fact. In Eq.~(\ref{eq:1-rv}) the first line corresponds to the interelectronic interaction between valence and hole particles taken with minus sign, the second and third lines are the one-electron corrections (self-energy, vacuum polarization, and counterterm) for the valence and hole particles, respectively. It is clear, that excitation of an electron from state $h$ to $v$ leads to the subtraction of an one-electron hole energy and addition of an one-electron valence energy, cf. Eq.~(\ref{eq:0-rv}). As one can see, the interelectronic interaction between valence (hole) particle with core electrons is not explicitly recognizable in Eq.~(\ref{eq:1-rv}). In fact, the SE and VP terms with the redefined vacuum propagator contain also the interelectronic interaction with the core electrons. As an example we show how the redefined vacuum expressions are linked to the usual ones for the valence SE and VP contributions. One has the following relationship \cite{Soguel_2021pra}
\begin{eqnarray}
\bra{v}  \Sigma_{\alpha}(\epsilon_{v}) \ket{v} &=& \bra{v}  \Sigma(\epsilon_{v}) \ket{v} - \sum_a I_{v a a v } (\Delta_{va}) \,, \nonumber \\
\bra{v} \Upsilon_{\alpha} \ket{v} &=& \bra{v} \Upsilon \ket{v} + \sum_a I_{v a v a} (0) \,,
\end{eqnarray}
where $\Sigma$ and $\Upsilon$ operators differ from corresponding $\alpha$-subscript-operators just by setting in Eqs.~(\ref{eq:se}) and (\ref{eq:vp}) $E^{F}_\alpha = 0$, respectively. In other words, to extract the interelectronic interactions arising from the one-particle graphs in the redefined vacuum Eq.~(\ref{eq:1-part}) one has to subtract the identical graph in the standard vacuum, as inferred from the above equations. Obviously the two-particle contribution Eq.~(\ref{eq:2-part}) is not affected by such manipulations. The resulting first order energy correction in the usual vacuum can be written as 
\begin{eqnarray}
\Delta E_{vh}^{(1)} &=&
\sum_a \left[I_{v a v a} (0) - I_{v a a v } (\Delta_{va})\right]\nonumber\\
&-&\sum_a \left[I_{h a h a} (0) - I_{h a a h } (\Delta_{ha}) \right]\nonumber\\
&+&F_{v_1 h_1} F_{v_2 h_2} \left[ I_{v_1 h_2 h_1 v_2} (\Delta_{h v}) - I_{v_1 h_2 v_2 h_1} (0)\right]\nonumber\\
&+&\bra{v} \Sigma(\epsilon_{v}) \ket{v}  +\bra{v} \Upsilon \ket{v} -U_{v v}\nonumber\\
&-&\bra{h} \Sigma(\epsilon_{h}) \ket{h} - \bra{h} \Upsilon \ket{h} + U_{h h}\,.
\label{eq:1-uv}
\end{eqnarray}
Now, the interaction between valence and hole particles with core electrons appears in the first and second lines of Eq.~(\ref{eq:1-uv}), respectively. Furthermore, the interelectronic interaction obtained (first three lines in Eq.~(\ref{eq:1-uv})) is in perfect agreement, when the Breit approximation is applied in the Coulomb gauge, with the one found in Ref.~\cite{PhysRevA.51.297}, where MBPT corrections up to the second order to the valence-hole state were derived. Other contributions (fourth and fifth lines) found are the SE and VP corrections to each particle, as expected.
 
Last but not least, the employment of a vacuum redefinition allows us to identify gauge invariant subsets in the usual vacuum. Following the logic and proofs presented in Ref.~\cite{Soguel_2021pra}, the contributions in the first, second, and third lines of Eq.~(\ref{eq:1-uv}) are gauge invariant independently. Moreover, each term in the fourth and fifth lines are also separately gauge invariant.
%
%
\section{Discussion and Conclusion}
\label{section:discussion}
The formalism and expressions presented in this work are ready to be applied to a closed shell atom or ion, such as Be-, Ne-like ions, etc. Be-like ions having the least number of electrons were already treated within the complete BSQED description. In particular, the ground-state and ionization energies were evaluated in Refs.~\cite{malyshev:2014:062517, malyshev:2015:012514}, respectively, while the transition energies between low-lying levels have been just recently addressed in Ref.~\cite{malyshev:2021:183001}. In last case, \cite{malyshev:2021:183001} the formal expressions have been derived with the TTGF method employing the redefined vacuum prescription, where the $1s^2$ shell has been considered as belonging to the vacuum and other two electrons have been treated as the valence electrons. For Ne-like ions such decomposition is rather complicated since one has to consider eight electrons to be the valence ones.

The energy levels in Ne-like ions have been investigated for a long time within MBPT approach \cite{avgoustoglou1992many, PhysRevA.51.297, Avgoustoglou1995, Avgoustoglou1996}. The discrepancies between theory and experiment for different transitions and ions, in absolute value, was less than 2 eV in the earliest work, which shrinks down to less than 1 eV in a more recent one \cite{Safronova2001}. In the case of Ne-like Ge a recent study \cite{PhysRevA.100.032516} reports the remarkable agreement up to $10^{-4}$ relative uncertainty between MBPT calculations and measured values. The QED effects have been incorporated at the first order via the Model Lamb-shift-operator approach \cite{shabaev:2013:012513, shabaev2015qedmod} and the interelectronic interaction has been captured with the Breit interaction treated in the vanishing frequency limit. Results have been compared with previous computations performed by Safronova {\it et al.} \cite{Safronova2001} for Ne-like Mo and showed less than 0.25 eV difference, in absolute value, except for one case. Another recent measurement campaign on Ne-like Eu \cite{Beiersdorfer2020} showed an agreement, in absolute value, of the order of 1 eV between MBPT predictions and experimental values. Overall, it shows that  MBPT treatment captures reliably the essence of the energy difference in Ne-like ions. Nevertheless, a rigorous BSQED treatment of valence-hole excitation in closed shell system was lacking so far. BSQED contributions could fill the energy gap by including rigorously all first order corrections in $\alpha$. It can help to achieve an outstanding agreement among theory and experiment, which could be especially interesting in view of possible applications as optical atomic clocks with valence-hole transitions in B$^+$, Al$^+$, In$^+$, and Tl$^+$ ions \cite{Ludlow:2015}. Additional application might be seen in searching for an explanation of the disagreement in oscillator-strength ratio in Ne-like Fe \cite{Kuhn:2020_Ne_iron, Ne-like_Fe:2022}.

To conclude, in the present paper we derived the two-time Green function suited for a valence-hole excitation in a closed shell system within the rigorous BSQED framework. The derivation was carried out in the redefined vacuum framework, which allow us to focus only on the particles which make the difference between the configurations. The complete first-order corrections were considered, the explicit formulas were derived, and the gauge invariant subsets were identified. Our results can readily be applied for the rigorous BSQED calculations of the transition energies in a closed shell system.
%

\section{Acknowledgments}
The theoretical investigations presented in Sections II and III were supported by Bundesministerium für Bildung und Forschung (BMBF) through project 05P21SJFAA  and in Section IV by the Russian Science Foundation (Grant No. 22-12-00258).
A.V.V. acknowledges financial support by the Government of the Russian Federation through the ITMO Fellowship and Professorship Program. 

\appendix
\begin{widetext}
\section{Zeroth-order Green function}
\label{appendix_B}
Here, the zeroth-order Green function of Eq.~(\ref{Green_zero}) is calculated explicitly. The reason for previously added terms and the role of prefactors role will become clearer. Let's start by expressing the matrix element of the zero-order Green function as
\begin{equation}
\bra{(vh)_{JM}} g_{\alpha}^{(0)} (E)\ket{(vh)_{JM}} = F_{vh} F_{vh} \bra{\alpha} b_h a_v \left[\frac{1}{2!}\int d^3 {\bf x}_1 d^3 {\bf x}_2 d^3 {\bf y}_1 d^3 {\bf y}_2 \mathcal{S}(E; {\bf x }_1, {\bf x}_2, {\bf y }_1, {\bf y }_2)  \right] a_v^{\dagger} b_h^{\dagger} \ket{\alpha} \,,
\end{equation}
where
\begin{eqnarray}
\mathcal{S}(E; {\bf x }_1, {\bf x}_2, {\bf y }_1, {\bf y }_2) &=&  \sum_{i,l>E_{\alpha}^F, j,k<E_{\alpha}^F}  a_i^{\dagger} a_l b_k^{\dagger} b_j \left[ \phi^{\dagger}_i( {\bf x }_1 ) \phi^{\dagger}_j( {\bf x}_2 ) \mathcal{G}_{\alpha}^{(0)} (E; {\bf x }_1, {\bf x}_2, {\bf y }_1, {\bf y }_2) \gamma^{0}_1 \gamma^{0}_2 \phi_k({\bf y }_1 ) \phi_l(  {\bf y }_2 ) \right] \nonumber \\
&+& \sum_{j,k>E_{\alpha}^F, i,l<E_{\alpha}^F}  a_j^{\dagger} a_k b_l^{\dagger} b_i \left[ \phi^{\dagger}_i( {\bf x }_1 ) \phi^{\dagger}_j( {\bf x}_2 ) \mathcal{G}_{\alpha}^{(0)} (E; {\bf x }_1, {\bf x}_2, {\bf y }_1, {\bf y }_2) \gamma^{0}_1 \gamma^{0}_2  \phi_k({\bf y }_1 ) \phi_l(  {\bf y }_2 ) \right] \nonumber \\
&-& \sum_{i,k>E_{\alpha}^F, j,l<E_{\alpha}^F}  a_i^{\dagger} a_k b_l^{\dagger} b_j \left[ \phi^{\dagger}_i( {\bf x }_1 ) \phi^{\dagger}_j( {\bf x}_2 ) \mathcal{G}_{\alpha}^{(0)} (E; {\bf x }_1, {\bf x}_2, {\bf y }_1, {\bf y }_2) \gamma^{0}_1 \gamma^{0}_2  \phi_k({\bf y }_1 ) \phi_l(  {\bf y }_2 ) \right] \nonumber \\
&-& \sum_{j,l>E_{\alpha}^F, i,k<E_{\alpha}^F}  a_j^{\dagger} a_l b_k^{\dagger} b_i \left[ \phi^{\dagger}_i( {\bf x }_1 ) \phi^{\dagger}_j( {\bf x}_2 ) \mathcal{G}_{\alpha}^{(0)} (E; {\bf x }_1, {\bf x}_2, {\bf y }_1, {\bf y }_2)  \gamma^{0}_1 \gamma^{0}_2  \phi_k({\bf y }_1 ) \phi_l(  {\bf y }_2 ) \right] \,.
\end{eqnarray}
The zeroth-order spectral representation of the Green function is given by
\begin{equation}
\mathcal{G}_{\alpha}^{(0)} (E; {\bf x }_1, {\bf x}_2, {\bf y }_1, {\bf y }_2)  = \frac{1}{2!} \frac{\mathcal{A}_{vh}^{(0)} ( {\bf x }_1, {\bf x}_2, {\bf y }_1, {\bf y }_2) }{E - E_{vh}^{(0)} + i\varepsilon} + \text{regular terms at } E \sim E_{vh}^{(0)}
\end{equation}
with 
\begin{eqnarray}
\mathcal{A}_{vh}^{(0)}({\bf x }_1, {\bf x}_2, {\bf y }_1, {\bf y }_2) &=& \bra{\alpha} \left[ \psi^{(0)}_{\alpha}(0,{\bf x}_1)  \bar{\psi}^{(0)}_{\alpha}(0, {\bf y}_1) \ket{(vh)_{JM}} \bra{(vh)_{JM}} \psi^{(0)}_{\alpha}(0, {\bf x}_2) \bar{\psi}^{(0)}_{\alpha}(0,{\bf y}_2) \right. \nonumber \\
&-& \left.  \psi^{(0)}_{\alpha}(0,{\bf x}_2) \bar{\psi}^{(0)}_{\alpha}(0, {\bf y}_1) \ket{(vh)_{JM}} \bra{(vh)_{JM}} \psi^{(0)}_{\alpha}(0, {\bf x}_1) \bar{\psi}^{(0)}_{\alpha}(0, {\bf y}_2)  \right. \nonumber \\
&-& \left. \psi^{(0)}_{\alpha}(0,{\bf x}_1) \bar{\psi}^{(0)}_{\alpha}(0, {\bf y}_2) \ket{(vh)_{JM}} \bra{(vh)_{JM}} \psi^{(0)}_{\alpha}(0, {\bf x}_2) \bar{\psi}^{(0)}_{\alpha}(0, {\bf y}_1)  \right. \nonumber \\ 
&+& \left. \psi^{(0)}_{\alpha}(0,{\bf x}_2) \bar{\psi}^{(0)}_{\alpha}(0, {\bf y}_2) \ket{(vh)_{JM}} \bra{(vh)_{JM}} \psi^{(0)}_{\alpha}(0, {\bf x}_1) \bar{\psi}^{(0)}_{\alpha}(0, {\bf y}_1)  \right] \ket{\alpha}\,.
\label{a_term}
\end{eqnarray}
As a first step the previous expression, $\mathcal{A}_{vh}^{(0)}({\bf x }_1, {\bf x}_2, {\bf y }_1, {\bf y }_2)$, can be further evaluated. Recalling that $a_i \ket{\alpha} = 0$ and $b_j \ket{\alpha} = 0$, $\bar{\psi} = \psi^{\dagger} \gamma^{0}$ and $F_{vh} F_{v h} = 1$ one easily gets 
\begin{equation}
    \mathcal{A}_{vh}^{(0)}  ( {\bf x }_1, {\bf x}_2, {\bf y }_1, {\bf y }_2) =  \left(  \phi_v ({\bf x}_1)  \phi_h ({\bf x}_2) - \phi_h ({\bf x}_1)  \phi_v ({\bf x}_2) \right) \left(  \bar{\phi}_h ({\bf y}_1)  \bar{\phi}_v ({\bf y}_2) - \bar{\phi}_v ({\bf y}_1)  \bar{\phi}_h ({\bf y}_2) \right) \,. 
\end{equation}
Two comments should be made at this point. If one were to consider Eq.~(\ref{Green_initial}), only the first term from the expanded above expression would be found. Furthermore in Eq.~(\ref{Green}), a factor one over two factorial was introduced. The reason is to get a proper normalized two-body wavefunction in the non-interacting case as it will be seen in the following. Using Eq.~(\ref{G_operator}), one contracts the operators to obtain
\begin{eqnarray}
\bra{(vh)_{JM}} g_{\alpha}^{(0)} (E)\ket{(vh)_{JM}}  &=&
\frac{1}{2!}\left [g_{\alpha,v_1 h_2 h_1 v_2}^{(0)}(E) - g_{\alpha,v_1 h_2 v_2 h_1}^{(0)}(E) + g_{\alpha,h_2 v_1 v_2 h_1}^{(0)}(E) - g_{\alpha,h_2 v_1 h_1 v_2}^{(0)}(E) \right ]\,,
\end{eqnarray} 
which can be re-written according to Eq.~(\ref{G_matrix_element}) as
\begin{eqnarray}
\bra{(vh)_{JM}} g_{\alpha}^{(0)} (E)\ket{(vh)_{JM}}  &=&
\frac{1}{E - E_{vh}^{(0)} + i\varepsilon} 
\nonumber\\ 
&\times&
\int d^3 {\bf x}_1 d^3 {\bf x}_2 \frac{1}{2!}
\left[\phi_v^{\dagger} ({\bf x}_1)  \phi_h^{\dagger} ({\bf x}_2) - \phi_h^{\dagger} ({\bf x}_1)  \phi_v^{\dagger} ({\bf x}_2) \right] \left[  \phi_v ({\bf x}_1)  \phi_h ({\bf x}_2) - \phi_h ({\bf x}_1)  \phi_v ({\bf x}_2) \right]
\nonumber\\ 
&\times& \int  d^3 {\bf y}_1 d^3 {\bf y}_2 \frac{1}{2!}
\left[  \phi_h^\dagger ({\bf y}_1)  \phi_v^\dagger ({\bf y}_2) - \phi_v^\dagger ({\bf y}_1)  \phi_h^\dagger ({\bf y}_2) \right] 
\left[  \phi_h ({\bf y}_1)  \phi_v ({\bf y}_2) - \phi_v ({\bf y}_1)  \phi_h ({\bf y}_2) \right] \nonumber \\
&=& \frac{1}{E - E_{vh}^{0} + i\varepsilon} \,.\,\, \Box
\end{eqnarray} 

\end{widetext}

%

\bibliography{bibliography}
\end{document}